\documentclass[aps,pra,twocolumn,groupedaddress]{revtex4}

\usepackage{graphicx}
\usepackage{amsmath}
\usepackage{amssymb}

\newcommand{\br}{{\bf r}}

\begin{document}
\bibliographystyle{prsty}

\title{Spin textures in condensates with large dipole moments}
\author{J.~A.~M.\ Huhtam\"aki,$^{1,2}$ M.~Takahashi,$^{1}$ T.~P.~Simula,$^{1,3}$ T.~Mizushima,$^{1}$ and K.~Machida$^{1}$}

\affiliation{$^1$Department of Physics, Okayama University, Okayama
700-8530, Japan} \affiliation{$^2$Department of Applied
Physics/COMP, Aalto University School of Science and Technology,
P.O. Box 15100, FI-00076 AALTO, Finland} \affiliation{$^3$School of
Physics, Monash University, Victoria 3800, Australia}

\begin{abstract}
We have solved numerically the ground states of a Bose-Einstein
condensate in the presence of dipolar interparticle forces using a
semiclassical approach. Our motivation is to model, in particular,
the spontaneous spin textures emerging in quantum gases with large
dipole moments, such as ${}^{52} {\rm Cr}$ or ${\rm Dy}$
condensates, or ultracold gases consisting of polar molecules.
For a pancake-shaped harmonic (optical) potential, we
present the ground state phase diagram spanned by the strength of
the nonlinear coupling and dipolar interactions. In an elongated
harmonic potential, we observe a novel helical spin texture. The
textures calculated according to the semiclassical model in
the absence of external polarizing fields are predominantly analogous
to previously reported results for a ferromagnetic $F=1$ spinor
Bose-Einstein condensate, suggesting that the spin textures arising
from the dipolar forces are largely independent of the value of the
quantum \mbox{number $F$} or the origin of the dipolar interactions.
\end{abstract}


\maketitle

\section{Introduction}

Long-range interparticle forces in a quantum system with a large
coherence length is an intriguing combination bound to exhibit a
host of fascinating phenomena. Perhaps the most timely example of
such a system is the gaseous atomic Bose-Einstein condensate (BEC)
subject to magnetic dipole-dipole forces~\cite{Lahaye2009}.

The dipolar interaction potential, decreasing as $r^{-3}$ in terms
of the interparticle distance $r$, dominates on length scales
determined by the coherence length. Other two-body interactions
present in the system, such as induced dipolar forces (van der
Waals), weaken typically much faster ($r^{-6}$) and become
negligible already over distances of an average interparticle
separation. A further interesting aspect of the dipole-dipole
interaction is its anisotropy enriching the already diverse
finite-size effects in trapped ultracold atomic gases. The magnetic
dipolar interaction in condensates has been predicted to give rise
to phenomena ranging from spin textures and spontaneous mass
currents~\cite{Takahashi2007,yi:020401,kawaguchi2006scg} to roton
minimum in the excitation spectrum~\cite{ODell2003,Santos2003},
linking the field into the study of liquid He II.

The realization of ${}^{52} {\rm Cr}$ condensates has provided means
of probing dipolar effects experimentally due to the exceptionally
large magnetic moments of the atoms~\cite{PhysRevLett.94.160401}.
The ground states of a chromium condensate have been studied
extensively~\cite{Goral2000,santos:190404,diener:190405,Makela2007,He2009}.
Anisotropic deformation of an expanding chromium condensate due to
dipolar forces has been observed~\cite{PhysRevLett.95.150406}, and
dipole-induced spin relaxation in an initially polarized ${}^{52}
{\rm Cr}$ has been linked to the famous Einstein-de Haas effect in
ferromagnets~\cite{santos:190404,kawaguchi:080405}. Also, collapse
and subsequent $d$-wave symmetric explosion of dipolar condensates
have been recently studied in the case of ${}^{52}{\rm Cr}$ both
experimentally and theoretically~\cite{Lahaye2008}. Chromium
condensates have been recently produced through optical
methods~\cite{Beaufils2008}.

The strength of the magnetic dipolar interaction is determined by
the atomic magnetic moment $\mu_M$ through the coupling constant
$g'_d=\mu_0 \mu_M^2 /4\pi$, where $\mu_0$ is the permeability of
vacuum. For example, for alkali condensates with total angular
momentum quantum number $F=1$, the magnetic moment is given by
$\mu_M=\mu_B g_F$, where $\mu_B$ is the Bohr magneton and
$g_F=1/2$ the Land\'e $g$-factor. Such systems are subject to weak dipolar
interactions, e.g., $g'_d/g' \sim 10^{-3}$ for ${}^{87} {\rm Rb}$,
where $g'=4\pi \hbar^2 \left( a_0 + 2 a_2 \right)/3m$ is the
mean-field density-density coupling constant. Here $a_0$ and $a_2$
are the $s$-wave scattering lengths in the channels with total spin
$0$ and $2$, and $m$ is the atomic mass. Nevertheless, dipolar
effects have been predicted to be observable in $F=1$ alkali BECs
even in the presence of a magnetic field~\cite{kawaguchi:110406},
which was recently confirmed experimentally based on time-evolution
study of a helical spin texture~\cite{Vengalattore2008}. It has also
been proposed that spin echo in spinor BECs could be utilized in
revealing dipole-dipole interactions~\cite{Yasunaga2008}.

The spontaneous occurrence of novel ground-state spin textures in
the absence of external magnetic fields requires typically stronger dipolar interactions,
$g'_d/g' \sim 10^{-2}$--$10^{-1}$. Hence, the ${}^{52}{\rm Cr}$
condensates consisting of particles with magnetic moments of $6\,
\mu_B$, as opposed to maximal magnetic moments of $1\, \mu_B$ in
alkali gases, seem more favorable for observing such effects.
Moreover, the rare-earth-metal element $\rm{Er}$, with a magnetic
moment of $7\, \mu_B$, has been cooled down to $\mu \rm{K}$
temperatures~\cite{Berglund2008}. Also, recent developments
in trapping and cooling of $\rm{Dy}$ with the largest atomic
magnetic moment of $10\, \mu_B$ yields a promising candidate for
observing the predicted spin textures~\cite{Lu2010}. Developments in
the study of ultracold polar molecules provides means of
investigating dipolar effects with large electric
moments~\cite{Bethlem1999,mancini:133203,ni2008hps}.

The study of alkali condensates based on a quantum mechanical
mean-field treatment predict spin textures with the smallest
possible value for the total angular momentum quantum number with
internal degrees of freedom, namely $F=1$. It is worthwhile to
approach an analogous problem from the other extreme limit
by treating the magnetic moments of the gas
classically~\cite{Takahashi2007,Takahashi2009}. By comparing the
results predicted by the two models, one may expect that if the
predictions agree, they could be of universal character for all
dipolar condensates and independent of the particular value of the
quantum number $F$. In general, the quantum mechanical order
parameter has $2F+1$ components and the short-range interaction term
contains $F+1$ independent coupling constants. Hence, it would be
very cumbersome to treat each value of $F$ separately with the
complexity of the problem increasing along with $F$.

Our semiclassical model is briefly described in Sec.~II. The order
parameter is written in an alternative form compared to previous
studies~\cite{Takahashi2007,Takahashi2009} in order to simplify
analysis and to increase numerical efficiency. The main results are
explained in Sec.~III: The ground states of the system in harmonic
traps of various geometries are described and the collapse of the
spin vortex state is analyzed briefly. The novel spin helix state is
introduced before concluding remarks of Sec.~IV.

\section{Model}
In this Section, we construct a phenomenological mean-field model
describing a trapped Bose-Einstein condensate with local as well as
nonlocal interparticle interactions. The model is equivalent to the
semiclassical approach previously studied in
\cite{Takahashi2007,Takahashi2009}, with the exception that now the
order parameter field is written in a cartesian basis yielding a set
of three Gross-Pitaevskii type of equations leading to more
efficient numerics.

The local interaction is assumed to be of the standard $s$-wave
form, with coupling constant $g'=4\pi \hbar^2 a/m$, where $a$ is the
$s$-wave scattering length. Henceforth, we will refer to its
dimensionless form $g=4\pi N a /a_r$, expressed in natural trapping
units: $\hbar \omega_r$ is the unit of energy with $\omega_r$ being
the radial trapping frequency of the confining harmonic potential,
and the radial harmonic oscillator length
$a_r=\sqrt{\hbar/m\omega_r}$ is the unit of distance. The number of
confined atoms is denoted by $N$. The nonlocal interaction is the
anisotropic dipole-dipole interaction with the dimensionless
coupling constant $g_d$ which is quantified in relation to $g$
throughout the article. Regardless of whether the origin of the dipolar
interactions is considered to be magnetic or electric,
we adopt notation and terminology as if it were of the former.
For the trapped system to be stable, we find
that for strong enough contact interaction, the value of $g_d$
should not exceed $\sim g/4$, a value very close to the number
calculated in the $F=1$ case~\cite{yi:020401}.

The order parameter is taken to be a three-component real-valued
vector $\psi=\left( \psi_x, \psi_y, \psi_z \right)$. It is
straightforward to show that all line defects in such
order-parameter space are topologically unstable, because any closed
curve on a sphere can be continuously transformed into a point which
corresponds to a spin polarized state~\cite{Mermin1979}.
Nevertheless, the energetically stable states can have nontrivial
spin textures.

In the present model, the particle density is assumed to be related
to the order parameter through $n(\br) = \sum_k \psi_k^2(\br)$ and
is normalized to unity, $\int d\br n(\br) = 1$. Let us make the
assumption that the system is ferromagnetic, and hence we may
require that all spins are pointing into the same direction within a
small enough region of space. With this simplification, the
magnitude of the magnetization density is related to the particle
density through
\begin{equation}
|{\bf M}(\br)| = \mu_M n(\br) = \mu_M \sum_k \psi_k^2(\br) =
\sqrt{\sum_k M_k^2(\br)},
\end{equation}
where $\mu_M$ is the magnetic moment of a single particle, and $M_k$
are the components of magnetization. By squaring, we obtain $\sum_k
M_k^2(\br) = \mu_M^2 \sum_k \psi_k^2(\br) n(\br)$, which is
satisfied if we define
\begin{equation}
M_k(\br) = \mu_M \psi_k(\br) \sqrt{n(\br)},
\end{equation}
relating the magnetization density to the order parameter. In the
following, we omit writing the constant $\mu_M$ explicitly and
assume it to be included in the coupling constant $g_d$.

The energy functional $E_{\rm tot}[\psi_x,\psi_y,\psi_z]$ can thus be written as
\begin{equation}
\small \label{FreeEnergy} E_{\rm tot} = \int \sum_k \psi_k
\hat{h} \psi_k d\br + \frac{g}{2} \int n^2 d\br +\frac{g_d}{2} \int \int
D(\br,\br') d\br d\br',
\end{equation}
where $\hat{h}=-\frac{1}{2} \nabla^2 + V_{\rm trap}(\br)$ is the
single-particle Hamiltonian and $V_{\rm trap}= \frac{1}{2}
\left(x^2+y^2+ \lambda^2 z^2\right)$ is the external trapping
potential expressed in natural trapping units. For now we will omit
external rotation and mass currents in the system, and hence the
kinetic energy is merely due to quantum pressure. The second term in
Eq.~(\ref{FreeEnergy}) describes the local mean-field $s$-wave
interaction with the coupling constant $g$, and the final term the
nonlocal dipole-dipole interaction
with
\begin{equation}
\small
D(\br,\br')=\left[{\bf M}(\br) \cdot {\bf M}(\br')-3\left({\bf M}(\br) \cdot {\bf e}_R \right) \left({\bf M}(\br') \cdot {\bf e}_R \right)\right]/R^3,
\end{equation}
where ${\bf R}=\br-\br'$ is the relative coordinate and ${\bf e}_R$ the unit vector along it.

Stationary states of the condensate are obtained by differentiating the energy functional with respect to the components of the order parameter with
the particle number constraint taken into account through a Lagrange multiplier $\mu$. Differentiation with respect to $\psi_j$ results in a set of three
Gross-Pitaevskii equations
\begin{equation}
\label{GP}
\small
\hat{h}\psi_j + gn \psi_j+\frac{g_d}{2} \left[\frac{\sum_k M_k I_k}{n}\psi_j+\sqrt{n}I_j \right]=\mu \psi_j.
\end{equation}
Here the functions $I_j$ are defined by
\begin{eqnarray}
\small
I_j(\br)=\int \left[ M_j(\br')-3 e_R^j \sum_l M_l(\br') e_R^l \right]/R^3 d\br',
\end{eqnarray}
with $e_R^l$ being the $l$th component of ${\bf e}_R$. These integrals may be further broken into convolutions. By applying the convolution theorem, we obtain
\begin{eqnarray}
\label{Integral_j} \small I_j(\br)= \mathcal{F}^{-1} \left[\sum_l
\mathcal{F}[M_l({\bf r})] \hat{f}_{lj}({\bf k}) \right],
\end{eqnarray}
where $\mathcal{F}$ stands for Fourier transform and
\mbox{$\hat{f}_{lj}({\bf k})=-\frac{4\pi}{3} \left( \delta_{lj}-3
k_l k_j/k^2 \right)$} is the Fourier transform of \mbox{$f_{lj}({\bf
r})=\left( \delta_{lj}-3 r_l r_j/r^2 \right)/r^3$}. The Fourier transforms are
efficiently evaluated by using Fast Fourier Transform.

From the general form of the GP equations, Eq.~(\ref{GP}), it is
possible to conclude that the spin-polarized texture is not a
stationary state in a confined three-dimensional system in the
absence of external polarizing fields when $g_d \ne 0$. Namely,
Eq.~(\ref{GP}) is of the form $A_{jk} \psi_k = b_j$, where
$b_j=\frac{g_d}{2}\sqrt{n}I_j \ne 0$, in general. In the
spin-polarized state, we may choose, say, the $z$-axis along the
polarization, whence $\psi_x=\psi_y=0$ yielding $b_x=b_y=0$ from the
general form above. When $g_d \ne 0$, this can be satisfied in
regions of non-vanishing density only if $I_x=I_y=0$. Hence, the
bracketed expression in Eq.~(\ref{Integral_j}) must vanish
identically. As $\hat{f}_{zx}$ and $\hat{f}_{zy}$ are non-vanishing
in any finite volume $d^3k$, continuity of $\psi_z$ implies
$M_z(\br)=0$, which is a contradiction. Such conclusion can also be
drawn from the quantum mechanical model by following similar
arguments.

Apart from the quantum pressure term in Eq.~(\ref{FreeEnergy}), the
spin model described above can be viewed as resulting from a
classical energy functional. However, the present model can also be
argued from the quantum mechanical spin-$F$ model constrained within
the ferromagnetic manifold~\cite{Ho1996}. With maximally aligned
spins, the order parameter at a fixed point ${\bf r}$ is of the form
$\psi = \sqrt{n} e^{i\theta} e^{-i \hat{F}_z \alpha/\hbar} e^{-i
\hat{F}_y \beta/\hbar} e^{-i \hat{F}_z \gamma/\hbar} |z \rangle =
\sqrt{n} e^{i(\theta-F \gamma)} e^{-i \hat{F}_z \alpha/\hbar} e^{-i
\hat{F}_y \beta/\hbar} |z \rangle$, where $\hat{F}_\alpha$ are the
hyperfine spin operators, and $\hat{F}_z |z\rangle = \hbar F
|z\rangle$. The order parameter of the classical spin model is
obtained if we neglect the phase factor $e^{i(\theta-F\gamma)}$ and
replace the quantum mechanical rotation operators by the classical
equivalents and the eigenstate $|z\rangle$ by the unit vector
pointing along the $z$-axis. Such substitution should be valid when
quantum fluctuations of the spin operator $\hat{\bf F}$ become
negligible. The relative fluctuations in the state $|z \rangle$ are
given by $\langle (\hat{\bf F} - \hbar F )^2\rangle/\langle \hat{\bf
F}^2 \rangle = 1/(F+1)$, which vanish in the limit of large $F$.
Possible mass currents arising from local spin-gauge symmetry are
neglected when $e^{i(\theta-F\gamma)}$ is set to unity, and the
kinetic energy reduces merely to the quantum pressure term in
Eq.~(\ref{FreeEnergy}).

\section{Results}
We have solved the ground states of the system with various values
of the coupling constants $g$ and $g_d$, and the aspect ratio
$\lambda$. Special emphasis is given to the pancake- and
cigar-shaped systems, for which we choose $\lambda=10$ and
$\lambda=0.10$--$0.50$, respectively.

\begin{figure}[!ht]
\includegraphics[width=250pt]{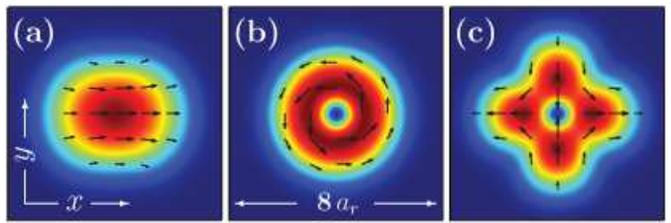}
\caption{\label{FIG1} (Color online) Magnetization ${\bf M}(\br)$
(arrows) and density $n(\br)$ (color) of the (a) flare and (b) spin
vortex states for $g=100$, $g_d/g=0.15$ in a trap with aspect ratio
$\lambda=10$. Both quantities are shown in the $z=0$ plane, the
density being nearly gaussian in the axial direction and $M_z$
small. For the chosen parameter values, the flare state in (a) is
the energetically favored configuration. Panel (c) illustrates a
spin vortex state with opposite spin winding compared to (b). Such
state is not energetically favorable for the parameter values
considered in this work. Each panel has dimensions $8\, a_r \times
8\, a_r$.}
\end{figure}

\subsection{Ground states in the pancake-shaped limit}
Let us first consider the case of a cylindrically symmetric harmonic
trap with strong, $\lambda=10$, confinement in the axial ($z$)
direction. In the presence of dipolar interactions, $g_d > 0$, the
magnetic moments tend to lie predominantly in the plane
perpendicular to the axial direction in order for the system to
minimize dipolar interaction energy.

For small enough value of $g_d/g$, the spin texture has typically
the flare structure which has been studied previously using the
semiclassical approach as well as the quantum mechanical mean-field
model in the $F=1$
case~\cite{Takahashi2007,yi:020401,kawaguchi2006scg}. Such state is
illustrated in Fig.~\ref{FIG1}(a) for, $g=100$ and $g_d/g=0.15$. The
arrows denote the local direction of magnetization ${\bf M}(\br)$,
whereas the color refers to the particle density $n(\br)$. The
repulsive interaction between parallel spins separated by a vector
perpendicular to the spin vectors causes the magnetization to
deviate from the polarized texture. The structure may also be
thought of as resulting from the presence of two spin vortices
located at the periphery of the cloud. The spin texture is
flare-like also in the $x$--$z$ -plane ($y=0$) due to finite $M_z$,
which is in accordance with the picture that a single toroidal spin
vortex encircles the cloud. In the flare state, the magnetization
has even parity.

\begin{figure}[!ht]
\vspace{10pt}
\hspace{-20pt}
\includegraphics[width=230pt]{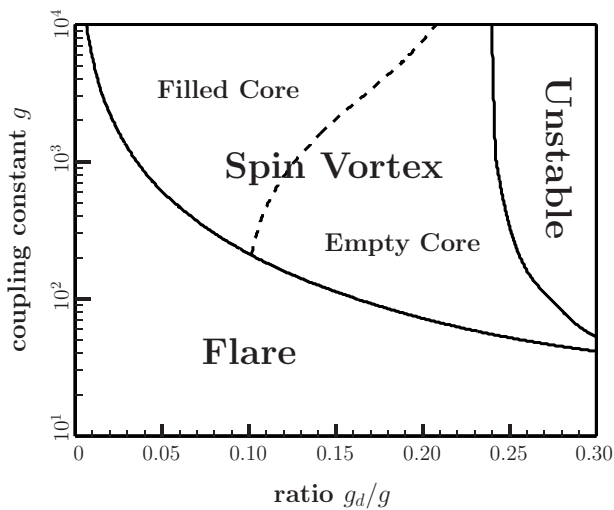}
\caption{\label{FIG2} Ground-state phase diagram of a dipolar
condensate in a harmonic trap with aspect ratio $\lambda=10$. The
effective contact interaction coupling constant $g$ is represented
in logarithmic scale on the vertical axis. The horizontal axis,
measuring the strength of dipolar interactions through the ratio
$g_d/g$, has linear scale. The phase diagram is divided into three
regions: flare (Fig.~\ref{FIG1}(a)), spin vortex
(Fig.~\ref{FIG1}(b)), and the region where the spin vortex becomes
unstable against collapse.}
\end{figure}

When the strength of dipolar interactions is increased, the ground
state undergoes a second order phase transition into a state hosting
a single spin vortex which is illustrated in Fig.~\ref{FIG1}(b) for
$g=100$ and $g_d/g=0.15$. The density is typically suppressed at the
core of the vortex. The spin vortex state has also been studied
previously within the semiclassical as well as the $F=1$
case~\cite{Takahashi2007,yi:020401,kawaguchi2006scg}. Analogously to
the flare state, the presence of the spin vortex results in a
texture which favors dipolar interactions by reducing the repulsive
interactions of parallel spins separated by a vector perpendicular
to their magnetization.
For example, close to the phase transition line in Fig.~\ref{FIG2}
with $g=1000$ and $g_d/g=0.05$, the differences in the kinetic,
potential, contact interaction, and dipolar energies of the flare
and the spin vortex states are $\Delta E_{\rm kin}=-0.16$, $\Delta
E_{\rm pot}=0.077$, $\Delta E_{\rm nl}=-0.11$, and $\Delta E_{\rm
dip}=0.28$, respectively, leading to a gain in total energy of
$\Delta E_{\rm tot}=0.077$ in units of $\hbar \omega_r$ per
particle.

Figure \ref{FIG1}(c) illustrates a spin vortex state with opposite
spin winding. For this texture, the angle between local
magnetization and the $x$--axis decreases as the vortex core is
circled around in the counterclockwise direction, whereas for the
state in Fig.~\ref{FIG1}(b), the angle increases. Such state is
found only as an excited solution in the present work. In a
larger dipolar system, one can construct energetically low-lying
spin vortex lattices by arranging the vortices presented in
Figs.~\ref{FIG1}(b) and \ref{FIG1}(c), and their negative
counterparts (${\bf M} \longrightarrow -{\bf M}$) in an alternating
square lattice. Both spin vortices presented in Fig.~\ref{FIG1} have
odd parity.

The ground-state phase diagram in the $(g,g_d/g)$--parameter plane
is shown in Fig.~\ref{FIG2}. The axes of the plane are chosen such
that the abscissa is proportional to the particle number $N$ and the
ordinate is independent of $N$ and proportional to the (bare)
dipolar coupling constant $g'_d$. The spin vortex state is
energetically favored for strong contact and dipolar interactions.
The flare state dominates the phase diagram in the limit of weak
contact interaction, $g \lesssim 50$, regardless of the strength of
dipolar interactions. The phase transition point from flare to spin
vortex state depends strongly on the value of $g$. The spin vortex
state becomes unstable towards collapse beyond the critical value of
$g_d/g \approx 0.25$ which depends only weakly on the value of $g$
for $g \gtrsim 500$.

With strong enough contact interaction $(g \gtrsim 1000)$ and weak
dipolar interaction, the cores of the spin vortices are filled with
particles whose magnetic moments are pointing in the axial
direction. The condensate gains trapping as well as contact
interaction energy by filling the vortex core.
The dashed line in the phase diagram of Fig.~\ref{FIG2} separates
the states with filled cores from states with empty cores. On the
line $\int n(r=0,z) dz/{\rm max}{\int n(r,z) dz}=0.01$ with
$r=\sqrt{x^2+y^2}$, whereas the ratio is close to unity in the upper
left corner. The finite axial magnetic moment due to the filled core
breaks the inversion symmetry of the state. Instead, the components
of ${\bf M}$ have the following symmetry: $\hat{P}_z
M_{x,y}=-M_{x,y}$, $\hat{P}_z M_z=M_z$, where $\hat{P}_z$ inverts
the sign of the $x$-- and $y$--coordinates keeping $z$ intact.

For large enough $g$, the flare state develops continuously into a
state with two spin vortices which have ferromagnetic cores as the
strength of dipolar interactions is increased. The magnetic moments
of the cores are pointing either into the same or opposite
directions, the two states being nearly degenerate irrespective of
the relative orientation. States hosting multiple spin vortices are
found to be energetically unfavorable compared to single spin vortex
states for the parameter values considered in this work.

The radial size of the spin vortex state diminishes significantly as
the strength of dipolar interactions is increased. This suggests
that the reason why the system becomes unstable at some critical
value of $g_d/g$ could be due to inward collapse of the condensate.
Local and global collapse of a dipolar condensate has been recently
studied numerically~\cite{Parker2009}.

In order to understand why the spin vortex solution ceases to exist
above the critical point, it is instructive to study scaling
transformations of the form
\begin{equation}
\label{ScalingTransformation} \hat{T}^{\sigma}(\tau) \psi_k(r,z) =
c^\sigma(\tau) \psi_k([1+\tau] r,[1+\sigma \tau] z),
\end{equation}
where $(r,z)$ are the cylindrical coordinates, $\tau$ is the scaling
parameter, $\sigma$ determines the ratio between axial and radial
scaling, and $c^\sigma(\tau)$ is chosen to ensure particle number
conservation.

Close to the critical point of collapse, the spin vortex state in a
pancake-shaped trap is the ground state of the system, and hence
lies in a minimum of the energy functional. Under transformations of
the form given in Eq.~(\ref{ScalingTransformation}), the total
energy becomes a function of the scaling parameter $\tau$, $E_{\rm
tot}^{\sigma}(\tau)=E_{\rm tot}[\hat{T}^{\sigma}(\tau)
\psi_k(r,z)]$. Deviation from the ground state always leads to
increase in energy, and hence the second derivative of the total
energy with respect to any one-parameter transformation must be
positive, $\partial^2_{\tau} E^{\sigma}_{\rm tot}(\tau)
\big|_{\tau=0}>0$. The existence of a transformation for which this
quantity vanishes indicates that the state becomes unstable against
such variation.

\begin{figure}[!ht]
\vspace{10pt}
\hspace{-20pt}
\includegraphics[width=230pt]{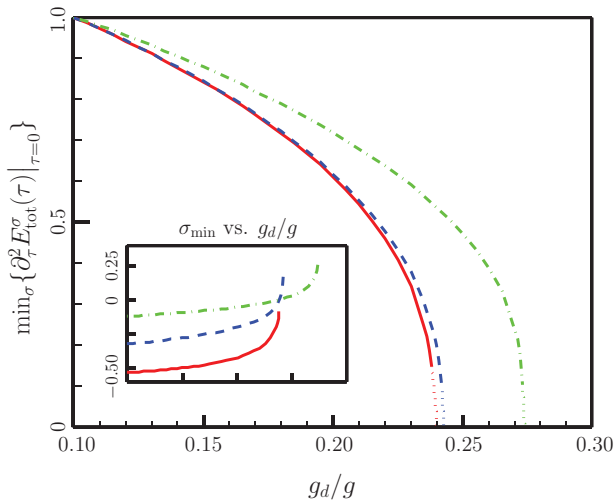}
\caption{\label{FIG3} (Color online) Second derivative of the total
energy with respect to a scaling transformation of the form given in
Eq.~(\ref{ScalingTransformation}) as a function of $g_d/g$ shown in
units of the corresponding quantity at $g_d/g=0.10$. The curves
correspond to the parameter values $g=10^4$ (solid), $g=10^3$
(dashed), and $g=10^2$ (dash-dotted). The value of $g_d/g$ for which
the second derivative vanishes indicates the critical strength
beyond which the spin vortex state becomes unstable against
collapse. The inset shows the ratio of axial and radial scaling for
which the minimal value of the bracketed expression in the main
figure is obtained.}
\end{figure}

Figure~\ref{FIG3} shows the value of
$\mathop{\rm min}_{\sigma} \{\partial^2_{\tau} E^{\sigma}_{\rm
tot}(\tau) \big|_{\tau=0} \}$
as a function of $g_d/g \in [0.10,0.30]$ scaled by the value at
$g_d/g=0.10$. The solid curve is for $g=10^4$, dashed for $g=10^3$,
and dash-dotted for $g=10^2$. The curves are extrapolated (dotted
lines) using the last few points to obtain an estimate for the
critical value for which the minimum in the energy functional
vanishes. The critical values are $g_d/g=(0.240,0.243,0.274)$,
respectively. The inset in Fig.~\ref{FIG3} depicts the value of
$\sigma$ for which $\partial^2_{\tau} E^{\sigma}_{\rm tot}(\tau)
\big|_{\tau=0}$
is minimized for each $g_d/g$, the horizontal axis being the same as
in the main graph. In the vicinity of the critical point for $g
\lesssim 10^3$, $\sigma>0$, showing that the collapsing cloud
shrinks both in radial and axial directions.

Based on a spinor $F=1$ study, a critical value of $g_d/g \approx
0.24$ has been previously reported for the existence of the spin
vortex state~\cite{yi:020401}, where the parameters are chosen such
that $g \approx 7000$ in the present study. As the current model is
expected to be accurate, apart from possible mass currents,
in the limit of large (classical) magnetic
moments, this agreement suggests that
the critical value for the collapse is universal and independent of
$F$.

\subsection{Ground states in the cigar-shaped limit}
Let us now consider solutions to Eq.~(\ref{GP}) in an elongated
trapping geometry with $\lambda \in [0.10, 0.50]$. For definiteness,
we will fix $g=10^4$ which corresponds to a number of $N \approx 1.5
\times 10^5$ ${}^{87} {\rm Rb}$ atoms in a harmonic trap with radial
frequency $\omega_r = 2\pi \times 100$ Hz. Regardless of the aspect
ratio $\lambda$, the solutions are found to exist only within the
interval $g_d/g \in [0,0.235]$, agreeing with the result previously
reported in the $F=1$ study~\cite{yi:020401}.

\begin{figure}[!ht]
\includegraphics[width=250pt]{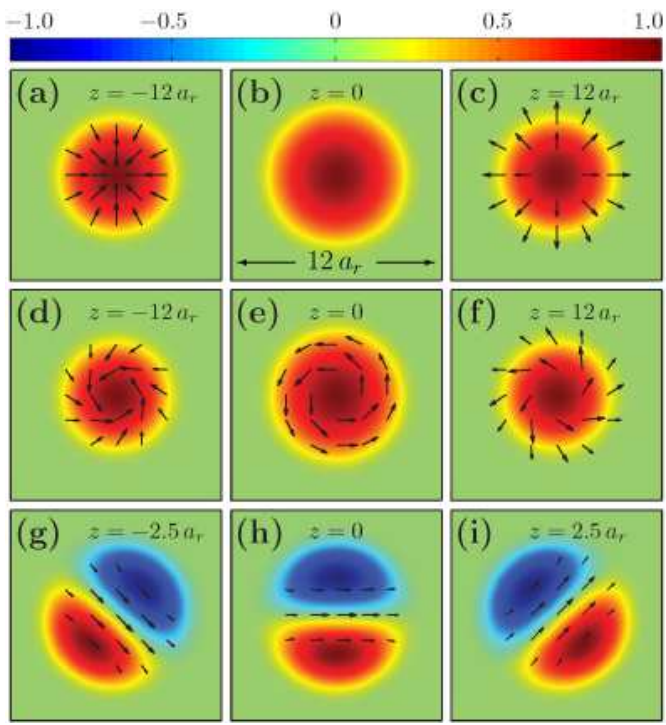}
\caption{\label{FIG4} (Color online) Spin textures in the flare,
(a)--(c), spin vortex, (d)--(f), and spin helix, (g)--(i), states in
a cigar-shaped trapping geometry with aspect ratio $\lambda=0.20$
and dipolar interaction strengths $g_d/g = 0.030,$ $0.080$, and
$0.030$, respectively. The arrows illustrate the magnetization
within a given radial cross-section projected onto the $x$--$y$
-plane, whereas the color refers to the axial magnetization $M_z$
normalized with respect to maximal magnetization within each state
separately. Each panel has dimensions $12\, a_r \times 12\, a_r$.}
\end{figure}

The spins tend to lie predominantly along the axial direction for
finite but sufficiently weak dipolar interactions. This ground state
resembles the flare state in the pancake-shaped limit, and it has
been discussed previously both in $F=1$ condensates as well as using
the semiclassical model~\cite{kawaguchi2006scg,Takahashi2007}.
Figures~\ref{FIG4}(a)--(c) illustrate the spin textures in three
radial cross-sections of the condensate. Here $\lambda=0.20$,
$g_d/g=0.030$, and the cross-sections are taken at $z=-12\, a_r, 0$,
and $12\, a_r$, respectively. The color depicts the $z$-component of
magnetization, $M_z({\br})$, and the color bar is scaled with
respect to the maximum magnetization, ${\rm max} |{\bf M}({\bf r})|$,
in the corresponding state. The arrows show the texture
projected onto the $x$--$y$ -plane (henceforth referred to as the
planar texture), with the length of the arrows scaled within each
panel separately, except in Fig.~\ref{FIG4}(b), for which $M_x=M_y=0$
by symmetry. The stability range of the flare state depends strongly
on the aspect ratio: For example, with $\lambda=0.50$, the flare
state is the ground state for $0<g_d/g \lesssim 0.01$, with
$\lambda=0.20$ for $0<g_d/g \lesssim 0.08$, whereas with
$\lambda=0.10$ the flare state dominates the entire stability
window.

Typically, for stronger dipolar interactions, a spin vortex texture
appears in the central region of the condensate, illustrated in
Figs.~\ref{FIG4}(d)--(f) for $\lambda=0.20$ and $g_d/g=0.080$. As
shown in Fig.~\ref{FIG4}(e), the planar texture near the center
resembles the spin vortex texture in the pancake-shaped geometry
discussed before, see Fig.~\ref{FIG1}(b). As one moves further away
from the center, the planar texture deforms continuously towards the
flare texture discussed in the previous paragraph, as depicted in
Figs.~\ref{FIG4}(d) and \ref{FIG4}(f). Although the magnetization in
both the flare and the spin vortex states has the same symmetry,
$\hat{P}_z M_{x,y}=-M_{x,y}$, $\hat{P}_z M_z=M_z$, the phase
transition is sharp, as illustrated below in Fig.~\ref{FIG5}(d).

In order to characterize the spin vortex state more precisely, we
define the following quantities: The axial {\it column density}
reads
\begin{equation}
n_z(z)=\int n({\bf r}) dx dy.
\end{equation}
This measures the number of atoms per unit length in the axial
direction and is normalized to unity. The average {\it twisting
angle} is given by
\begin{equation}
\alpha(z)=\left \langle \arccos \left[ \frac{\hat{{\bf r}}_{xy}
\cdot {\bf M}_{xy}({\bf r})}{M_{xy}({\bf r})} \right] \right
\rangle,
\end{equation}
where $\hat{{\bf r}}_{xy}=\left(x {\bf e}_x+y {\bf e}_y
\right)/\sqrt{x^2+y^2}$ and the averaging is taken over vectors
${\bf M}_{xy}$ whose length exceeds $1\, \%$ of the maximum of
the planar magnetization $M_{xy}=\sqrt{M_x^2+M_y^2}$.
This quantity characterizes the twisting of
the magnetization in plane, yielding zero (or $\pi$) for the
flare-like textures, Figs.~\ref{FIG4}(a)--(c), and $\pi/2$ for the
spin vortex texture, Fig.~\ref{FIG4}(e). The average twisting angle
is essentially independent of the radial distribution of the
density. Finally, we define the average {\it tilting angle} through
\begin{equation}
\beta(z)=\left \langle \arctan \left[ \frac{M_{xy}({\bf r})}{M_{z}({\bf r})} \right] \right \rangle,
\end{equation}
where the averaging is evaluated as above. The tilting angle is
related to the pitch of the helical streamlines obtained by
following the local direction of magnetization in the spin vortex
state, {\it c.f.}~Ref.~\cite{yi:020401}.

\begin{figure}[!ht]
\includegraphics[width=240pt]{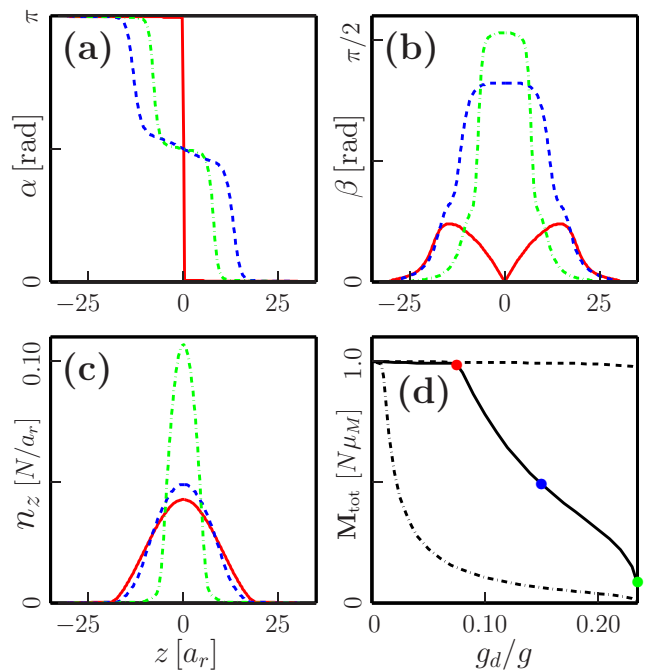}
\caption{\label{FIG5} (Color online) The twisting angle $\alpha(z)$
(a), tilting angle $\beta(z)$ (b), and the axial column density
$n_z(z)$ (c) for three values of the dipolar interaction strength
$g_d/g=(0.075, 0.15, 0.235)$ shown with solid, dashed, and
dash-dotted lines, respectively. The total magnetization ${\bf
M}_{\rm tot}$, which is directed along the $z$-axis by
symmetry, is shown in (d) for the aspect ratios $\lambda=0.10$
(dashed), $0.20$ (solid), and $0.50$ (dash-dotted). The dots in (d)
refer to the values of $g_d/g$ used in (a), (b), and (c).}
\end{figure}

\begin{figure*}[ht!]
\includegraphics[width=500pt]{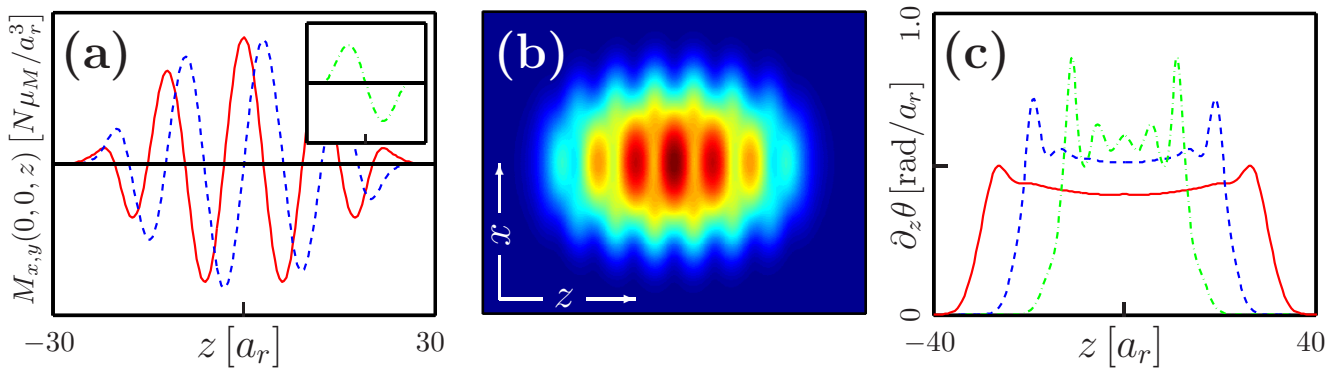}
\caption{\label{FIG6} (Color online) (a) $M_x$ (solid) and $M_y$
(dashed) on the $z$--axis as a function of $z$ in the spin helix
state for $\lambda=0.10$ and $g_d/g=0.20$. The dash-dotted curve in
the inset shows $M_z$ along the $y$--axis in the range $[-5,5]\,
a_r$. The vertical axis spans the interval $[-0.0028,0.0028]\, N
\mu_M/a_r^3$ both in the main figure and the inset. (b) The column
density $n(x,z)=\int n({\br}) dy$ illustrating density oscillations
characteristic of the spin helix state for strong dipolar
interactions. The parameters are as in (a), and the field of view is
$8\, a_r \times 60\, a_r$. (c) Wave vector of the spin helix state
for $g_d/g=(0.080, 0.20, 0.23)$ shown with solid, dashed and
dash-dotted curves, respectively. The peaks at the ends of the cloud
are finite-size effects.}
\end{figure*}

Figure~\ref{FIG5}(a) shows the twisting $\alpha(z)$ in the flare
(solid) and spin vortex state (dashed and dash-dotted) for the
dipolar interaction strengths $g_d/g = \left( 0.075, 0.15, 0.235
\right)$, respectively. The flare state is chosen from the
neighborhood of the transition point to a spin vortex state.
However, $\alpha(z)$ remains nearly zero (or $\pi$) over the whole
length of the cloud. Small deviation from zero shows that the flare
state has even parity only approximately. As the dipolar interaction
strength is increased, a spin vortex enters the system. Hence, the
twisting angle decreases continuously from $\pi$ to $0$ along the
length of the condensate. A plateau of $\alpha(z) \approx \pi/2$
forms in the central region of the system for strong dipolar
interactions, $g_d/g \gtrsim 0.15$. The width of the plateau
decreases for increasing $g_d/g$ due to shrinking of the cloud.

The tilting angle $\beta(z)$ is shown in Fig.~\ref{FIG5}(b) for the
same parameter values as in Fig.~\ref{FIG5}(a). It remains
relatively small in the flare state and experiences a sudden
increase at the center of the system when the ground state hosts a
spin vortex. For very strong dipolar interactions, $\beta(z) \approx
\pi/2$ in the central region, slightly even exceeding $\pi/2$ due to
interaction with the axial magnetization of the core region. The
small lobes in $\beta(z)$ close to the top and bottom of the cloud
are remainders of the flare state.

The strength of dipolar interactions affects the spatial density
profile of the spin vortex state significantly. The axial column
density $n_z(z)$ is shown in Fig.~\ref{FIG5}(c) for the parameter
values used in 5(a) and 5(b). Not only does the system shrink in the
radial, but also the axial direction with increasing $g_d/g$. Also,
the column density appears to be slightly bimodal for strong enough
dipolar interactions: the column density is enhanced in the central
region of the condensate where the spin vortex lies in order for the
system to gain dipolar energy. The width of the plateau in
Fig.~\ref{FIG5}(a) due to the presence of the spin vortex matches
the size of the central profile in the bimodal density distribution.
The bimodality appears more vividly in elongated systems with
$\lambda < 0.20$. In the extreme limit of $g_d/g=0.235$, the
density $n(\br)$ in the spin vortex state is significantly reduced
close to the center of the trap where $\br \approx 0$.

Figure \ref{FIG5}(d) depicts the total magnetization ${\bf M}_{\rm
tot}=\int {\bf M}(\br) d\br$ in the flare and spin vortex states as
a function of $g_d/g$. Due to symmetry, the total magnetization is
along the axial direction. The dashed, solid and dash-dotted lines
correspond to the aspect ratios $\lambda=(0.10, 0.20, 0.50)$,
respectively. The sudden drops in total magnetization indicate the
phase boundary between the flare and spin vortex states. Whereas the
flare state dominates the entire stability window for very elongated
trapping geometry ($\lambda=0.10$), the ground state in a prolate
system ($\lambda=0.50$) hosts a spin vortex already with
$g_d/g=0.01$. The phase transition points for different values of
$\lambda$ agree qualitatively with the analogous results in the
$F=1$ study~\cite{yi:020401}.

It is reasonable to expect that there exist also stationary states
with the opposite symmetry compared to the flare and spin vortex
states, i.e., $\hat{P}_z M_{x,y}({\br})=M_{x,y}({\br})$, $\hat{P}_z
M_z({\bf r})=-M_z({\bf r})$. There indeed exist low-energy solutions
to Eq.~(\ref{GP}) with such symmetry, to which we refer to as spin
helices. The spin helix state is found, e.g. with $\lambda=(0.10,
0.20, 0.50)$ in the entire stability interval $0 < g_d/g \le 0.235$
of the system. This state resembles closely the state studied in
Ref.~\cite{Vengalattore2008}, where the helical spin texture is
created by using a transient magnetic field gradient. According to
our simulations, the stationary spin helix state exists also in a
ferromagnetic $F=1$ system with dipolar interactions, which will be studied in more detail
elsewhere. Dynamical instability of a similar structure in the
absence of dipolar interactions has been studied
recently~\cite{Cherng2008}.

The spin helix texture is illustrated for $\lambda=0.20$ and
$g_d/g=0.030$ in Fig.~\ref{FIG4}(g)--\ref{FIG4}(i), where the radial
cross-sections are taken at $z=(-2.5\, a_r, 0, 2.5\, a_r)$,
respectively. On the $z$--axis, the magnetization lies in the
$x$--$y$ -plane, as a consequence of the antisymmetry of $M_z$.
Further away from the $z$--axis and perpendicular to the
magnetization on the axis, $M_z$ becomes the dominant component. The
whole planar texture rotates about the $z$--axis as a function of
the $z$--coordinate, traversing typically through several cycles
along the length of the condensate.

Energetically, the spin helix state appears to be favored by strong
dipolar interactions and not too elongated geometries. For example,
with $\lambda=0.50$, the helix becomes energetically favorable
compared to the spin vortex state between $0.050 < g_d/g < 0.10$. It
is challenging to pinpoint the exact location of the phase
transition point due to near degeneracy of the two states. Near the
critical value of $g_d/g = 0.235$, the difference in the total
energy between the spin vortex and helix states is $\Delta E_{\rm
tot} \approx 0.1\, \hbar \omega_r$ per particle in favor of the
helix, which is roughly $2\%$ of the total energy. For
$\lambda=0.20$, the helix state appears to be the minimal energy
texture only for $g_d/g \gtrsim 0.20$, whereas for $\lambda=0.10$,
the flare state lies $2\%$--$7\%$ lower in energy over the entire
stability range of the system.

The number of cycles in the helix texture increases as the
condensate is elongated, and thus for the sake of clarity we
illustrate it as an excited state for $\lambda=0.10$ and
$g_d/g=0.20$ in Figs.~\ref{FIG6}(a)--(c). In Fig.~\ref{FIG6}(a), the
solid and dashed curves show the components $M_x$ and $M_y$ along
the $z$--axis, respectively, and the $M_z$ component along the
$y$--axis is shown in the inset by the dash-dotted curve.

The spin helix can be thought of as two elongated stripes, polarized
along the $z$--axis in the opposite directions, intertwined around
one another. The helical texture on the $z$--axis arises due to
continuous twisting of the magnetization via the $x$--$y$ -plane. In
the $F=1$ case, quantized spin vortices of opposite winding
penetrate through the axially polarized ferromagnetic stripes,
forming an intertwined spin vortex pair. Intertwining of two mass
vortices has been previously studied in relation to the splitting of
a doubly quantized vortex in a scalar
condensate~\cite{PhysRevA.68.023611,Huhtamaki2006}.


Figure \ref{FIG6}(b) illustrates the column density $n(x,z)=\int
n({\br}) dy$ for the same state as in Fig.~\ref{FIG6}(a), red
denoting area of high and blue of vanishing particle density. With
strong dipolar interactions density oscillation appear spontaneously
due to the helix spin texture: for a fixed point in the $x$--$y$
-plane close to the surface region of the cloud, the axial
magnetization $M_z$ is an oscillating function of $z$. The particle
density is suppressed in the vicinity of the nodes of $M_z$ and
enhanced at the anti-nodes due to dipolar interactions.


As a measure of the pitch of the spin helix, we define the angle
\begin{equation}
\theta(z)=\arctan \left[ \frac{M_y(0,0,z)}{M_x(0,0,z)} \right].
\end{equation}
The derivative $\partial \theta/\partial z$ yields the wave vector
of the helix, which is plotted in Fig.~\ref{FIG6}(c) for
$\lambda=0.10$ and $g_d/g=(0.080,0.20,0.23)$ with the solid, dashed,
and dash-dotted curves, respectively. The wave vector tends to
increase for stronger dipolar interactions, which is reasonable
because the dipolar coherence length decreases as $\xi_{\rm d}
\propto g_d^{-1/2}$. The peaks in $\partial \theta / \partial z$ at
the top and bottom of the cloud are finite-size effects: The texture
may adjust freely into an energetically favorable configuration at
the edge as one of the boundary conditions due to continuity of the
order parameter is liberated. Oscillations penetrate along the whole
length of the condensate for $g_d/g=0.23$. These oscillations
enhance rapidly as the strength of dipolar interactions is increased
even further.

The number of cycles in the spin helix state decreases as the aspect
ratio $\lambda=\omega_z/\omega_r$ is increased, until in spherical
geometry, the direction of the spin on the $z$--axis twists only
through half a cycle along the length of the system, c.f.
Fig.~\ref{FIG6}(a). Interestingly, the spin vortex state, for which
the magnetization has the opposite symmetry with respect to
inversion about the $z$--axis, reduces to the spin helix state, rotated by
$(\pm)\pi/2$ about ${\bf e}_z \times {\bf M}_{\rm h}$, where ${\bf
M}_{\rm h}$ is the magnetization at the trap center in the helix
state.

\section{Summary and Conclusions}
We have studied spin textures arising from dipolar interactions in
gaseous Bose-Einstein condensates of particles with
large permanent dipole moments. The theory is based on a
semiclassical model treating the dipole moments of the bosons
classically.

The observed spin textures in clouds confined in harmonic trapping
potentials agree qualitatively with previously reported results for
an $F=1$ system~\cite{yi:020401,kawaguchi2006scg}, such as
${}^{87}{\rm Rb}$. Moreover, the ground-state phase transition
points with respect to the strength of dipolar interactions seem to
agree roughly both with weak and tight axial trapping frequency.
The qualitative agreement in the observations drawn from the two
models suggests that similar textures and phase diagrams are to be
expected also for ferromagnetic systems with $F>1$ and for condensates
consisting of electric dipoles.

A major difference between our semiclassical model and the quantum
mechanical model describing a magnetic system lies in the existence
of mass currents: In the quantum mechanical case, phase gradients of
the components of the order parameter emerge spontaneously possibly
giving rise to mass currents which are absent in the semiclassical
treatment, c.f.~the last paragraph of \mbox{Sec.~II.} The main
effect from taking spontaneous mass currents into account regarding
the present results would be that the phase transition line between
the flare and the spin vortex states in Fig.~\ref{FIG2} is shifted
to the right due to increased kinetic energy of the spin vortex
state. Spin dynamics of ferromagnetic condensates has been studied
recently in the long-wavelength limit using a hydrodynamic
model~\cite{Lamacraft2008}.

In addition to solving the ground states of the system for various
parameter values, we investigate the spin vortex state in the
extreme limit of the dipolar interaction strength $g_d/g \approx
1/4$. For larger dipolar interactions, the state becomes unstable
against collapse of the cloud due to strong attractive forces
overwhelming the quantum pressure term and repulsive interparticle
interactions. The estimated point of instability agrees well with
the value observed in the $F=1$ study~\cite{yi:020401}.

In the limit of tight axial confinement, two ground states are
observed, namely, the flare and the spin vortex states. For prolate
geometries, an additional spin helix texture appears as a low-energy
stationary state. The helix is the ground state of the system only
in slightly prolate condensates and for strong dipolar interactions. This
state is most likely related to the $S$ state reported
in~\cite{yi:020401}, and is especially interesting in relation to
the experimental observation of dipolar effects in ${}^{87}{\rm Rb}$
utilizing a similar spin texture~\cite{Vengalattore2008}. The
magnetization pattern of the spin helix gives rise to spontaneous
density oscillations in the stationary state for strong dipolar
interactions. As in the case of the spin vortex and flare textures,
the helix state ceases to exist for $g_d/g \gtrsim 1/4$.

\acknowledgements The authors would like to thank Japan Society for
the Promotion of Science (JSPS) for financial support.
Y.~Kawaguchi is acknowledged for useful comments.

\bibliography{JabRef}

\end{document}